\documentclass[twocolumn,aps,prl,showpacs]{revtex4}
\usepackage{amssymb}
\usepackage{graphicx}
\usepackage{times}
\usepackage{dcolumn}
\usepackage{bm}
\usepackage{epsfig}
\usepackage[english]{babel}
\usepackage{latexsym}
\usepackage{graphics}
\usepackage{subfigure}

\def\be{\begin{equation}}
\def\ee{\end{equation}}
\def\bea{\begin{eqnarray}}
\def\eea{\end{eqnarray}}
\def\bse{\begin{subequations}}
\def\ese{\end{subequations}}

\def\be{\begin{equation}}
\def\ee{\end{equation}}
\def\bea{\begin{eqnarray}}
\def\eea{\end{eqnarray}}
\def\bse{\begin{subequations}}
\def\ese{\end{subequations}}

\def\be{\begin{eqnarray}}
\def\ee{\end{eqnarray}}

\begin{document}

\title{Bell's inequality and universal quantum gates in a cold atom chiral
fermionic $p$-wave superfluid}
\author{Chuanwei Zhang, Sumanta Tewari, S. Das Sarma}
\affiliation{Condensed Matter Theory Center, Department of Physics, University of
Maryland, College Park, MD 20742}

\begin{abstract}
We propose and analyze a probabilistic scheme to entangle two spatially
separated topological qubits in a $p_{x}+ip_{y}$ superfluid using controlled
collisions between atoms in movable dipole traps and unpaired atoms inside
vortex cores in the superfluid. We discuss how to test the violation of
Bell's inequality with the generated entanglement. A set of universal
quantum gates is shown to be implementable \textit{deterministically} using
the entanglement despite the fact that the entangled states can only be
created probabilistically.
\end{abstract}

\maketitle

\emph{Introduction:} Topological quantum computation affords the amazing
possibility that qubits and quantum gates may be realized using only the
topological degrees of freedom of a system \cite{Kitaev}. Since these
degrees of freedom, by definition, are insensitive to weak local
perturbations, the resulting computational architecture should be free of
environmental decoherence, a major stumbling block to quantum computation.
In a class of topological systems, the requisite (non-Abelian) statistical
properties \cite{Read, Nayak} are provided by the presence of Majorana
fermion excitations described by the self-hermitian operators $\gamma
^{\dagger }=\gamma $. These excitations have been shown to occur naturally
at the cores of vortices in a 2D spinless $p_{x}+ip_{y}$ superfluid or
superconductor \cite{Green,Ivanov,Tewari}, where the interacting fermions
are described by the many body Pfaffian wavefunction \cite{Read}. (It seems
likely, but remains to be verified, that this wavefunction also describes
the essential physics of the filling fraction $\nu =5/2$ fractional quantum
Hall (FQH) system \cite{Read, Green}). It is encouraging that the spinless $%
p_{x}+ip_{y}$ superfluid of fermionic cold atoms is potentially realizable
in an optical trap tuned close to a $p$-wave Feshbach resonance \cite%
{Regal,Gurarie,Cheng}. Our current work establishes the possibility of
testing Bell's inequality in a cold atom $p$-wave fermionic superfluid on
the way to eventual universal topological quantum computation using vortices
in such a system.

In a $p_{x}+ip_{y}$ superfluid, one can define a topological qubit using a
group of four vortices. Since the states of the qubit are associated with
the composite states of the four \textit{spatially separated} Majorana
fermion excitations, they are immune to local environmental errors. One can
implement some single-qubit gates by adiabatically moving (braiding) one
vortex around another within the same vortex complex defining the qubit.
Since the associated unitary transformations are purely statistical, there
is, in principle, no error incurred in these gating operations. However, it
is well known \cite{Bravyi} that such a braid operation of one vortex from
one qubit around another from a different qubit fails to provide a two-qubit
gate: the topological braiding operations allowed in a $p_{x}+ip_{y}$
superfluid, as in its FQH Pfaffian counterpart, are not computationally
sufficient.

The principal reason why a $p_{x}+ip_{y}$ superfluid is not computationally
universal is that two qubits cannot be entangled using only the topological
braiding operations. Any composite state of the two qubits, accessible by
braiding one excitation around another, can always be written as a product
of the states of the individual qubits \cite{Bravyi}. Therefore, in light of
its experimental relevance, it is important to examine the problem of
creating quantum entanglement in a $p_{x}+ip_{y}$ superfluid via some other,
possibly non-topological, means (without incurring too much error) which,
coupled with the available braiding transformations, may lead to universal
quantum computation. This is all the more important because the other, more
exotic, non-Abelian topological states, e.g. the SU$\left( 2\right) $
Read-Rezayi state \cite{Rezayi}, which can support universal computation via
only the topologically protected operations \cite{Freedman,Bonesteel}, are
presently much beyond experimental reach. In the $5/2$ FQH state,
non-topological interference of charge-carrying quasiparticle currents along
different trajectories \cite{Bravyi,Walker,Sarma} was proposed to entangle
qubits. Such an approach is not suitable for the superfluid, because the
non-Abelian excitations here are vortices, which do not carry electric
charge.

In this Letter, we show how to entangle two spatially separated topological
vortex qubits in a cold atom $p_{x}+ip_{y}$ superfluid by using two other,
movable, external dipole traps. (The two-state systems formed by the atoms
in the movable, external traps will be referred as \textquotedblleft \textit{%
flying qubit}\textquotedblright ). Controlled cold collisions between an
atom in the dipole trap and an atom at the vortex qubit yield entanglement
between the flying qubit and the vortex qubit. Subsequently, a measurement
on a system comprising two flying qubits, entangled with two different
vortex qubits, collapses the two vortex qubits on an entangled state. We
show how to test the violation of Bell's inequality with the obtained
entanglement. Finally, we show how to \textit{deterministically} implement a
set of universal quantum gates using the entangled state, although the
entanglement among the vortex qubits itself can only be generated with a
50\% success probability. It is important to mention that the entanglement
can be generated and purified off-line, and so the non-topological nature of
the corresponding operations does not degrade the topological quantum
computation.

\begin{figure}[t]
\includegraphics[scale=0.39]{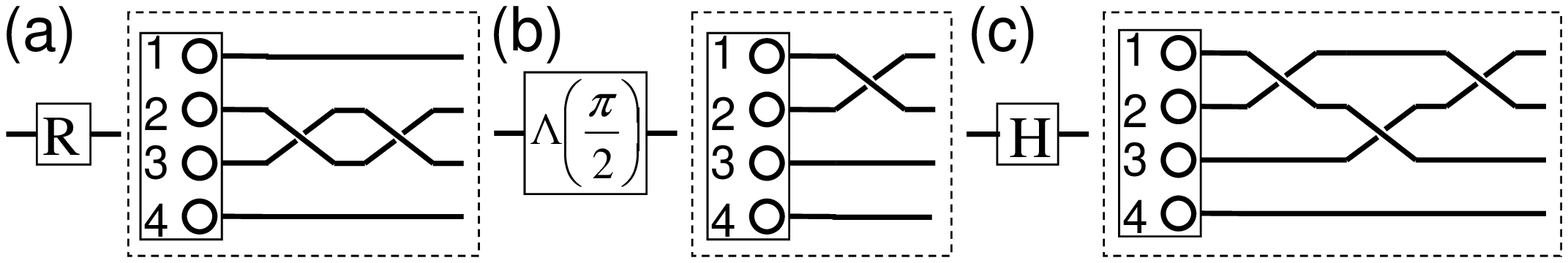} \vspace*{-0.3cm}
\caption{(a) Single qubit flip gate R $=-i\protect\sigma _{x}$. (b) Single
qubit phase gate $\Lambda \left( \protect\pi /2\right) =$ diag$\left(
1,i\right) $. (c) Hadamard gate H $=\frac{1}{\protect\sqrt{2}}\left(
\protect\begin{array}{cc}
1 & 1\protect \\
1 & -1%
\protect\end{array}%
\right) $.}
\label{rr1}
\end{figure}

\emph{Topological qubit and flying qubit:} Consider a quasi-two dimensional (%
$xy$ plane) $p_{x}+ip_{y}$ superfluid of spin-polarized atoms \cite%
{Regal,Gurarie,Cheng}, where vortices in the superfluid can be generated
through rotation or external laser fields. For each vortex, there exists a
zero energy state that supports a Majorana fermion mode $\gamma $ \cite%
{Read,Ivanov,Tewari}. Two Majorana fermion states in two vortices can be
combined to create an ordinary fermionic state $c=\left( \gamma _{1}+i\gamma
_{2}\right) /2$. Therefore, a natural definition of a vortex qubit may be
given in terms of the unoccupied, $\left\vert 0\right\rangle $, or occupied,
$\left\vert 1\right\rangle =c^{\dag }\left\vert 0\right\rangle $, states of
two Majorana vortices. Here the occupied state $\left\vert 1\right\rangle $
contains an unpaired Fermi atom\ inside the cores of the vortex pair
(\textquotedblleft unpaired\textquotedblright\ as opposed to
\textquotedblleft paired\textquotedblright\ in a cooper pair in the
superfluid). However, such a definition does not allow the superposition of
the basis states, i.e., the states, $(\left\vert 0\right\rangle \pm
\left\vert 1\right\rangle )/\sqrt{2}$, do not exist because they violate the
conservation of the total topological charge (the superfluid condensate
conserves the fermion number modulo 2). To overcome this difficulty, a
topological vortex qubit is defined through two pairs of vortices, i.e.,
with the states $\left\vert 0\right\rangle _{V}\equiv \left\vert
00\right\rangle $ (the two vortex pairs, (1,2) and (3,4), are both
unoccupied), and $\left\vert 1\right\rangle _{V}\equiv \left\vert
11\right\rangle $ (the two vortex pairs are both occupied). The
superposition states, $(\left\vert 0\right\rangle _{V}\pm \left\vert
1\right\rangle _{V})/\sqrt{2}$, are now allowed. Note also that these two
states do not mix, via any unitary braiding operations, with the other two
states of the four-vortex complex, $\left\vert 10\right\rangle $, $%
\left\vert 01\right\rangle $. Various intra- and inter-pair vortex braiding
operations within a single qubit give rise to various single-qubit gates
\cite{Walker,Georgiev} (e.g. qubit-flip gate R, phase gate $\Lambda (\pi /2)$
and the Hadamard gate H) as depicted schematically in Fig.~\ref{rr1}.
Finally, the state of the vortex qubit can be read out in the $\left\{
\left\vert 0\right\rangle _{V},\left\vert 1\right\rangle _{V}\right\} $
basis by fusing the vortices pairwise and detecting the number of unpaired
atoms in the core \cite{Tewari}.

The flying qubit is constructed using an atom trapped in the ground state of
a movable optical dipole trap which is itself formed by overlapping two
identical laser beam traps. One laser beam trap can then be adiabatically
moved out to split the composite trap into two traps, $L,R$, see Fig.~\ref%
{rr2}(a). This yields a superposition state for the atom, $\left( \left\vert
01\right\rangle _{LR}+\left\vert 10\right\rangle _{LR}\right) /\sqrt{2}$.
Here $L$ and $R$ denote the left and the right traps, respectively. Now,
concentrating on the left trap only, one can define the two states of the
flying qubit, $\left\vert 0\right\rangle _{F}=$ $\left\vert 01\right\rangle
_{LR}$, $\left\vert 1\right\rangle _{F}=$ $\left\vert 10\right\rangle _{LR}$%
. Note that the two states of the qubit are distinguished by the absence ($%
\left\vert 0\right\rangle _{F}$) and the presence ($\left\vert
1\right\rangle _{F}$) of the atom in the left dipole trap.

\begin{figure}[b]
\includegraphics[scale=0.55]{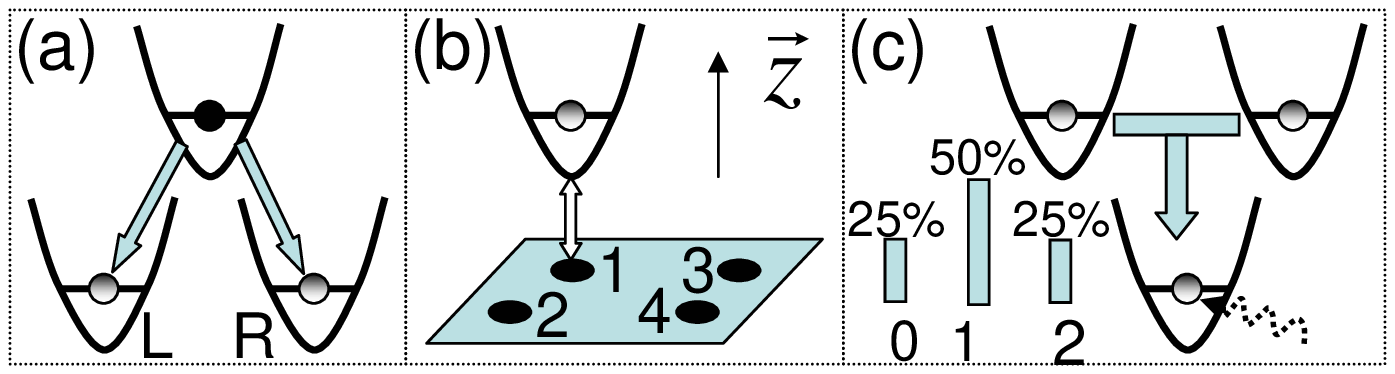}
\caption{(a) Construction of the flying qubit by splitting a composite
dipole trap into two traps. (b) Realization of the gate CP$\left( \protect%
\pi \right) $ by controlled collisions of atoms. (c) Two flying qubits are
merged into one and the number of atoms is measured through fluorescence
signals to create entanglement between two topological qubits (see text).}
\label{rr2}
\end{figure}

\emph{Entanglement between two topological qubits}: As is well known~\cite%
{Bravyi}, two topological qubits \textit{cannot} be entangled by braiding
one vortex from one qubit around another from the second qubit. However,
using the flying qubits as auxiliary degrees of freedom, one can generate
entangled states between the two qubits. The basic idea of the entanglement
generation is illustrated in Fig.~\ref{rr2}(b, c). Initially, a vortex
qubit, $V$, is prepared in the state $\left\vert 0\right\rangle _{V}$. A
Hadamard gate H is applied to the qubit that transfers the state to $%
\left\vert \phi \right\rangle _{V}=\left( \left\vert 0\right\rangle
_{V}+\left\vert 1\right\rangle _{V}\right) /\sqrt{2}$. By splitting a
composite dipole trap in two parts (Fig. \ref{rr2}(a)), the flying qubit $F$
is prepared in the state $\left\vert \psi \right\rangle _{F}=\left(
\left\vert 0\right\rangle _{F}+\left\vert 1\right\rangle _{F}\right) /\sqrt{2%
}$. The flying qubit is then moved near to the vortex qubit (Fig.~\ref{rr2}%
(b)) so that the trapped atom can collide with the unpaired fermi atom, if
any, in the vortex pair. As shown below, such a collision process yields a
controlled phase gate, CP$\left( \pi \right) \equiv \exp \left( i\pi
n_{V}n_{F}\right) $, between the flying qubit and the vortex qubit, where $%
n_{V}=0,1$ is the number of unpaired fermi atom in the vortex pair and $%
n_{F}=0,1$ is the number of atom in the flying qubit. It is easy to see that
the gate CP$\left( \pi \right) $ gives rise to the transformation,%
\[
\left\vert \psi \right\rangle _{F}\left\vert \phi \right\rangle
_{V}\rightarrow \left[ \left\vert 0\right\rangle _{F}\left( \left\vert
0\right\rangle _{V}+\left\vert 1\right\rangle _{V}\right) +\left\vert
1\right\rangle _{F}\left( \left\vert 0\right\rangle _{V}-\left\vert
1\right\rangle _{V}\right) \right] /2,
\]%
which can be transferred to an entangled state%
\begin{equation}
\left\vert \Phi \right\rangle _{FV}=\left( \left\vert 0\right\rangle
_{F}\left\vert 0\right\rangle _{V}+\left\vert 1\right\rangle _{F}\left\vert
1\right\rangle _{V}\right) /\sqrt{2}  \label{state1}
\end{equation}%
between the flying qubit and the vortex qubit by applying a Hadamard gate on
the vortex qubit.

Two vortex qubits can be entangled by a projection measurement on the flying
qubits of two entangled states $\left\vert \Phi \right\rangle _{F_{1}V_{1}}$
and $\left\vert \Phi \right\rangle _{F_{2}V_{2}}$. The dipole traps of the
two flying qubits are spatially merged and the atom number is measured
through fluorescence signals (Fig.~\ref{rr2}(c)). From the combined state,%
\begin{eqnarray}
\left\vert \Phi \right\rangle _{F_{1}V_{1}}\left\vert \Phi \right\rangle
_{F_{2}V_{2}} &=&\frac{1}{2}(\left\vert 00\right\rangle
_{F_{1}F_{2}}\left\vert 00\right\rangle _{V_{1}V_{2}}+\left\vert
11\right\rangle _{F_{1}F_{2}}\left\vert 11\right\rangle _{V_{1}V_{2}}
\nonumber \\
&+&\left\vert 01\right\rangle _{F_{1}F_{2}}\left\vert 01\right\rangle
_{V_{1}V_{2}}+\left\vert 10\right\rangle _{F_{1}F_{2}}\left\vert
10\right\rangle _{V_{1}V_{2}}),  \nonumber
\end{eqnarray}%
where $\left\vert 00\right\rangle _{F_{1}F_{2}}=\left\vert 0\right\rangle
_{F_{1}}\left\vert 0\right\rangle _{F_{2}}$ etc., it is easy to deduce the
probabilities for the three possible outcomes: one atom (50\%), zero atom
(25\%), two atoms (25\%). In the last two cases, the states of the vortex
qubits are projected to $\left\vert 0\right\rangle _{V1}\left\vert
0\right\rangle _{V_{2}}$ and $\left\vert 1\right\rangle _{V_{1}}\left\vert
1\right\rangle _{V_{2}}$, respectively, and are not entangled. Therefore, in
these cases the above procedure for creating the entangled states, $%
\left\vert \Phi \right\rangle _{F_{1}V_{1}}$ and $\left\vert \Phi
\right\rangle _{F_{2}V_{2}}$, need to be repeated. However, in the case
where the measurement produces one atom, the quantum state of the two qubits
is projected to the entangled state $\left( \left\vert 0\right\rangle
_{V1}\left\vert 1\right\rangle _{V_{2}}+\left\vert 1\right\rangle
_{V_{1}}\left\vert 0\right\rangle _{V_{2}}\right) /\sqrt{2}$, which can be
transferred to the expected entangled state%
\begin{equation}
\left\vert \Psi \right\rangle _{V_{1}V_{2}}=\left( \left\vert 0\right\rangle
_{V1}\left\vert 0\right\rangle _{V_{2}}+\left\vert 1\right\rangle
_{V_{1}}\left\vert 1\right\rangle _{V_{2}}\right) /\sqrt{2}  \label{state2}
\end{equation}%
using simple qubit-flip gates. Note that the above entangled state can only
be created with a 50\% success probability. For later use, the gate
representing the generation of entanglement is denoted as \textquotedblleft
EG\textquotedblright .

The remaining problem for the entanglement generation is how to realize the
controlled phase gate, CP$\left( \pi \right) $, between the flying qubit and
the vortex qubit. In Fig.\ref{rr2}b, the center of the dipole trap, $\mathbf{%
\vec{r}}_{0}(t)=z_{0}\left( t\right) \mathbf{\vec{e}}_{z}$ (with the core of
vortex 1 as origin) is adiabatically brought from a distance $d_{0}\mathbf{%
\vec{e}}_{z}$ above the $z=0$ plane, where the wavepackets of the dipole
trapped atom and the unpaired atom in the vortex pair (1,2) do not overlap,
to a distance zero, where they do. The collision phases between the atoms
are dynamic phases. They are different for different quantum states of
flying and vortex qubits with different total energy,%
\begin{equation}
E\left( i,j\right) =E_{F}\left( i\right) +E_{V}\left( j\right) +\Delta
E_{c}\left( i,j\right) ,  \label{energy}
\end{equation}%
where, $i,j=0,1$ correspond to the quantum states $\left\vert 0\right\rangle
$ and $\left\vert 1\right\rangle $, respectively. $E_{F}\left( 0\right) =0$
and $E_{F}\left( 1\right) =\int d^{3}\mathbf{r}\alpha ^{\ast }\left( \mathbf{%
\vec{r}-\vec{r}}_{0}\left( t\right) \right) $$\left[ -\hbar ^{2}\nabla
^{2}/2m_{F}+V_{F}\left( \mathbf{\vec{r}-\vec{r}}_{0}\left( t\right) \right) %
\right] $ $\alpha \left( \mathbf{\vec{r}-\vec{r}}_{0}\left( t\right) \right)
+E_{g}$ are the energies of the flying qubit in the states $\left\vert
0\right\rangle $ and $\left\vert 1\right\rangle $, respectively. $%
V_{F}\left( \mathbf{\vec{r}-\vec{r}}_{0}\left( t\right) \right) $ is the
harmonic potential of the dipole trap, and $\alpha \left( \mathbf{\vec{r}-%
\vec{r}}_{0}\left( t\right) \right) $ is the ground state wavefunction of
the dipole trapped atom with mass $m_{F}$. $E_{g}$ is the interaction energy
between the dipole trapped atom and the paired BCS condensate. The second
term $E_{V}\left( j\right) $ corresponds to the energy of the fermionic
state in the vortex cores near the dipole trap. Because these states are the
solutions of the Bogoliubov-de Gennes (BdG) equations with eigenvalue zero, $%
E_{V}\left( j\right) =0$ for $j=0,1$ \cite{Tewari}.

The last term describes the collision energy \cite{Zoller} between dipole
trapped atoms and unpaired fermi atoms, and \textit{is non-zero only if both
the flying qubit and the vortex qubit are in the occupied state,}%
\begin{equation}
\Delta E_{c}\left( 1,1\right) =\frac{g}{2}\int d^{3}\mathbf{r}\left\vert
\alpha \left( \mathbf{\vec{r}}-\mathbf{\vec{r}}_{0}\left( t\right) \right)
\right\vert ^{2}\delta n_{V}\left( \mathbf{\vec{r}}\right) .  \label{cene}
\end{equation}%
Here $g$ is the collision interaction strength and $\delta n_{V}\left(
\mathbf{\vec{r}}\right) $ denotes the changes of the atom density from the
BCS condensate density. Using the standard harmonic trap wavefunction for $%
\alpha \left( \mathbf{\vec{r}-\vec{r}}_{0}\left( t\right) \right) $ and the
wavefunction for the zero-energy mode obtained from the solution of the BdG
equations, we find%
\begin{equation}
\Delta E_{c}\left( 1,1\right) =\hbar \Omega _{\pm }\exp \left(
-z_{0}^{2}\left( t\right) /\bar{a}^{2}\right)  \label{cene2}
\end{equation}%
where $\bar{a}^{2}=a_{D}^{2}+a_{V}^{2}$, with $a_{D}$ and $a_{V}$ the
oscillation lengths for harmonic trapping potentials along the $z$ direction
of the dipole trap and the superfluid, respectively. $\hbar \Omega _{\pm }$
is the characteristic energy scale for the collision interaction, which is
determined by the collision interaction strength $g$ as well as the overlap
between the wavefunctions of the dipole trapped atoms and unpaired fermions
in the vortex pair (1,2). Note that the occupied zero energy fermionic state
of the vortex pair (1,2) is as likely to contain extra atoms (corresponds to
$\Omega _{+}$) as to miss atoms (corresponds to $\Omega _{-}$) compared to
the BCS condensate, which yields $\Omega _{+}=-\Omega _{-}$.

The state-dependent energy (\ref{energy}) yields a state-dependent dynamic
phase%
\begin{equation}
\varphi \left( i,j\right) =\varphi _{F}\left( i\right) +\phi _{c}\left(
i,j\right)  \label{phase}
\end{equation}%
where $\varphi _{F}\left( i\right) =\frac{1}{\hbar }\int_{-\tau }^{\tau
}E_{F}\left( i\right) dt$ is the single qubit phase that can be incorporated
in the definition of the flying qubit, the collision phase $\phi _{c}\left(
i,j\right) =\frac{1}{\hbar }\int_{-\tau }^{\tau }\Delta E_{c}\left(
i,j\right) dt$, and $\mp \tau $ denote the time when the dipole trap center $%
\mathbf{\vec{r}}_{0}\left( t\right) $ moves from and back to the initial
place $d_{0}\mathbf{\vec{e}}_{z}$. Assuming that $\mathbf{\vec{r}}_{0}\left(
t\right) $ varies adiabatically as $z_{0}\left( t\right) /d_{0}=\eta \left(
\exp \left( t^{2}/\tau _{r}^{2}\right) -1\right) /\left( 1+\eta \exp \left(
t^{2}/\tau _{r}^{2}\right) \right) $ with the parameter $\eta =\exp \left(
-\tau _{i}^{2}/\tau _{r}^{2}\right) $, the controlled collision phase can be
written as,%
\begin{equation}
\theta \equiv \phi _{c}\left( 1,1\right) =\Omega _{\pm }\tau _{r}\int_{-\bar{%
\tau}}^{\bar{\tau}}\exp \left[ -\Upsilon \eta \frac{e^{\bar{t}^{2}}-1}{%
1+\eta e^{\bar{t}^{2}}}\right] d\bar{t}  \label{phase2}
\end{equation}%
where $\Upsilon =d_{0}^{2}/\bar{a}^{2}$ and time in the above integration
has been scaled by $\tau _{r}$. With a set of parameters for $^{6}$Li, $%
a_{D}=a_{V}=0.4\mu m$, $d_{0}=10a_{D}=4\mu $m, $\tau _{r}=\tau
_{i}=3.57/\Omega $, $\tau =10\tau _{r}$, $s$-wave scattering length $%
a_{s}\sim 53$nm, and the vortex core size $\xi \sim 1\mu $m, we estimate $%
\Omega _{\pm }\sim \pm 2\pi \times 6.6$kHz, $\tau \sim 0.86$ms and the phase
$\theta =\pm \pi $ (i.e., $\exp \left( i\theta \right) =-1$). Therefore the
controlled phase gate CP$\left( \pi \right) $ can be realized.

\begin{figure}[b]
\includegraphics[scale=0.55]{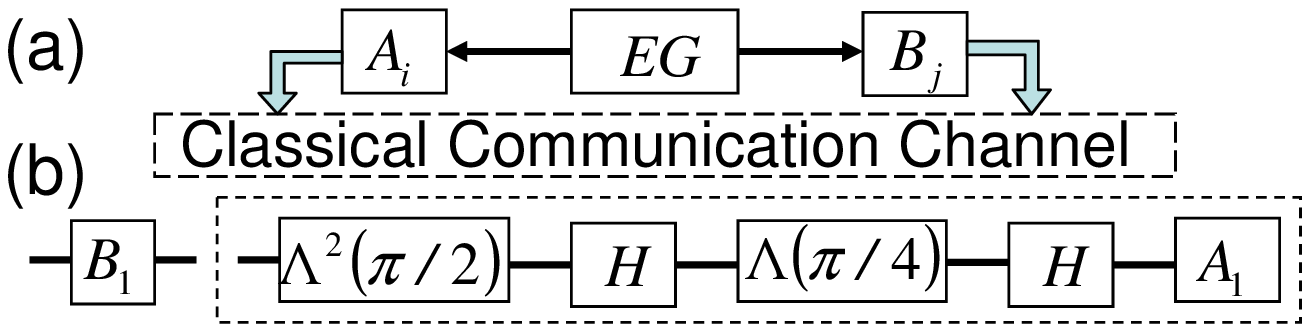}
\caption{(a) Testing the violation of the CHSH inequality. (b) The
realization of the B$_{1}$ measurement in (a).}
\label{rr3}
\end{figure}

\emph{Violation of Bell's inequality:} The entangled state $\left\vert \Psi
\right\rangle _{V_{1}V_{2}}$ between two remote vortex qubits can be used to
test the violation of the CHSH inequality, a variant of the Bell's
inequality \cite{Clauser}. Violation of the CHSH inequality would establish
the quantum non-locality between the two vortex qubits. A schematic diagram
of this test is given in Fig. \ref{rr3}. The test requires to measure the
vortex qubits along four different directions: $A_{1}=\sigma
_{z}^{V_{1}}\otimes I^{V_{2}}$, $A_{2}=\sigma _{x}^{V_{1}}\otimes I^{V_{2}}$%
, $B_{1}=-I^{V_{1}}\otimes \left( \sigma _{z}^{V_{2}}+\sigma
_{x}^{V_{2}}\right) /\sqrt{2}$, $B_{2}=I^{V_{1}}\otimes \left( \sigma
_{z}^{V_{2}}-\sigma _{x}^{V_{2}}\right) /\sqrt{2}$. After the measurements,
two parties at $V_{1}$ and $V_{2}$ need to communicate their results through
classical channel. After repeated measurements, the statistical average $%
L=\left\langle A_{1}B_{1}\right\rangle +\left\langle A_{2}B_{2}\right\rangle
+\left\langle A_{2}B_{1}\right\rangle -\left\langle A_{1}B_{2}\right\rangle $
is evaluated. The quantum non-locality of the entangled state yields $L=2%
\sqrt{2}$, which violates the CHSH inequality for local realism, $L\leq 2$
\cite{Clauser}. 

It is easy to convince oneself that the above four measurements correspond
to measuring the two vortex qubits in four different bases which are
eigenstates of the respective operators, $A_{1}:V_{1}$ on $\left\{
\left\vert 0\right\rangle _{V_{1}},\left\vert 1\right\rangle
_{V_{1}}\right\} $; $A_{2}$:$V_{1}$ on $\left\{ \left( \left\vert
0\right\rangle _{V_{1}}+\left\vert 1\right\rangle _{V_{1}}\right) /\sqrt{2}%
,\left( \left\vert 0\right\rangle _{V_{1}}-\left\vert 1\right\rangle
_{V_{1}}\right) /\sqrt{2}\right\} $; $B_{1}:V_{2}$ on $\left\{ a\left\vert
0\right\rangle _{V_{2}}+b\left\vert 1\right\rangle _{V_{2}},b\left\vert
0\right\rangle _{V_{2}}-a\left\vert 1\right\rangle _{V_{2}}\right\} $; $%
B_{2}:V_{2}$ on $\{{a\left\vert 0\right\rangle _{V_{2}}-b\left\vert
1\right\rangle _{V_{2}}},b\left\vert 0\right\rangle _{V_{2}}+a\left\vert
1\right\rangle _{V_{2}}\}$, where $a=\cos \left( \pi /8\right) $, $b=\sin
\left( \pi /8\right) $. In the experiment, $A_{1}$ is a fusion measurement
of the number of unpaired atoms in the vortices \cite{Tewari}. Measurements $%
A_{2}$, $B_{1}$, and $B_{2}$ can be implemented by first applying suitable
single-qubit operations to the qubits to transfer their measurement bases to
$\left\{ \left\vert 0\right\rangle _{V},\left\vert 1\right\rangle
_{V}\right\} $, following by fusion measurement $A_{1}$. The corresponding
single-qubit operations are $A_{2}$: $H$; $B_{1}$: $H\Lambda \left( e^{i\pi
/4}\right) H\Lambda ^{2}\left( \pi /2\right) $; $B_{2}$: $H\Lambda \left(
e^{i\pi /4}\right) H\Lambda ^{2}\left( -\pi /2\right) $, where $\Lambda
\left( e^{i\pi /4}\right) =$ diag$\left( 1,e^{i\pi /4}\right) $ is a single
qubit phase gate. $\Lambda \left( e^{i\pi /4}\right) $ cannot be implemented
through topologically protected braiding operations and its realization is
discussed in the next section.

\emph{Universal quantum gates: } It is well known that a set of quantum
gates \cite{Bravyi,Walker}%
\begin{equation}
H,\text{ \ }\Lambda \left( e^{i\pi /4}\right) ,\text{ \ }\Lambda \left(
\sigma _{z}\right)  \label{gates}
\end{equation}%
are sufficient to simulate any quantum circuit, where $\Lambda \left( \sigma
_{z}\right) =$ diag$\left( 1,1,1,-1\right) $ is the two-qubit controlled
phase gate between two vortex qubits. Among these three gates, the Hadamard
gate $H$ can be implemented using the topological braiding operations. The
single-qubit phase gate $\Lambda \left( e^{i\pi /4}\right) $ can be realized
by bringing two vortices 1 and 2 of a qubit close together \cite{Walker}.
The tunneling between two vortices induces an energy splitting $\Delta
E_{z}\left( t\right) $ between the state $\left\vert 0\right\rangle _{V}$
and $\left\vert 1\right\rangle _{V}$, which yields a relative dynamic phase $%
\exp $ $\left( i\int_{-T_{p}}^{T_{p}}\Delta E_{z}\left( t\right) dt\right) $
between these two states, with $2T_{p}$ as the total tunneling period. A
properly chosen tunneling process with $\int_{-T_{p}}^{T_{p}}\Delta
E_{z}\left( t\right) dt=\pi /4$ yields the gate $\Lambda \left( e^{i\pi
/4}\right) $.

A controlled phase gate $\Lambda \left( \sigma _{z}\right) $ between two
arbitrary vortex qubits can be realized \textit{deterministically} provided
one has been able to create the entangled state $\left\vert \Psi
\right\rangle $ between two vortex qubits. Considering two vortex qubits G
and Q (with the constituent vortices G$_{1}$, G$_{2}$, Q$_{1}$, Q$_{2}$
etc.), we note that $\Lambda _{GQ}\left( \sigma _{z}\right) =\Lambda
_{G}\left( \pi /2\right) \Lambda _{Q}\left( \pi /2\right) \exp \left( i\pi
\gamma _{G_{1}}\gamma _{G_{2}}\gamma _{Q_{1}}\gamma _{Q_{2}}/4\right) $,
where the last term involves interaction among four vortices. The
requirement of a four-vortex interaction is indeed the reason why the
two-qubit gate cannot be implemented using braiding operations which can
lead to only two-vortex (statistical) interactions. The four-vortex operator
can be implemented using one additional vortex pair $(\gamma _{W_{1}},\gamma
_{W_{2}})$ (initially prepared in state $\left\vert 0\right\rangle $) by
noting that \cite{Kitaev2},%
\begin{equation}
\exp \left( i\pi \gamma _{G_{1}}\gamma _{G_{2}}\gamma _{Q_{1}}\gamma
_{Q_{2}}/4\right) =2U_{\mu \nu }P_{\mu }^{\left( 2\right) }P_{\nu }^{\left(
4\right) },  \label{projection}
\end{equation}%
where $P_{\pm }^{\left( 2\right) }=(1\mp i\gamma _{Q_{1}}\gamma _{W_{1}})/2$
and $P_{\pm }^{\left( 4\right) }=(I\pm \gamma _{G_{1}}\gamma _{G_{2}}\gamma
_{Q_{2}}\gamma _{W_{1}})/2$ are non-destructive measurements which project
the state of the vortices to the eigenstates of the operators $-i\gamma
_{Q_{1}}\gamma _{W_{1}}$ and $\gamma _{G_{1}}\gamma _{G_{2}}\gamma
_{Q_{2}}\gamma _{W_{1}}$. $U_{\mu \nu }$ are corresponding braiding
operations for different measurement results $\{\mu \nu \}$, $%
U_{++}=U_{--}^{\dag }=e^{\pi \gamma _{Q_{1}}\gamma _{W_{2}}/4}$, $%
U_{+-}=i\Lambda _{G}\left( i\right) \Lambda _{Q}\left( i\right) e^{\pi
\gamma _{Q_{1}}\gamma _{W_{2}}/4}$, $U_{-+}=i\Lambda _{G}\left( i\right)
\Lambda _{Q}\left( i\right) e^{-\pi \gamma _{Q_{1}}\gamma _{W_{2}}/4}$. Here
$e^{\pi \gamma _{Q_{1}}\gamma _{W_{2}}/4}$ is just the exchange of the
vortices $\gamma _{Q_{1}}$ and $\gamma _{W_{2}}$.

$P_{\pm }^{\left( 2\right) }$ can be realized via a basis transformation
method. We exchange the vortices $\gamma _{Q_{1}}$ and $\gamma _{W_{1}}$ to
transfer two eigenstates of $-i\gamma _{Q_{1}}\gamma _{W_{1}}$ to $\left\{
\left\vert 00\right\rangle _{QW}\text{, }\left\vert 11\right\rangle
_{QW}\right\} $ or $\left\{ \left\vert 10\right\rangle _{QW}\text{, }%
\left\vert 01\right\rangle _{QW}\right\} $, depending on the total
topological charge of the four vortices $\gamma _{Q_{1}}$, $\gamma _{Q_{2}%
\text{, }}\gamma _{W_{1}}$ and $\gamma _{W_{2}}$. We then apply a fusion
measurement on the vortex pair $(\gamma _{W_{1}}$, $\gamma _{W_{2}})$ to
determine whether the state is $\left\vert 0\right\rangle _{W}$ or $%
\left\vert 1\right\rangle _{W}$, which correspond to the eigenvalues $+1$ or
$-1$ of the projection measurements $P_{\pm }^{\left( 2\right) }$. After the
fusion measurement, the vortex pair $(\gamma _{W_{1}}$, $\gamma _{W_{2}})$
is recreated in the state $\left\vert 0\right\rangle _{W}$. If the result of
the fusion measurement is the state $\left\vert 1\right\rangle _{W}$, this
state is recovered by applying a single-qubit flip operator R. Vortices $%
\gamma _{Q_{1}}$ and $\gamma _{W_{1}}$ are exchanged again to transfer the
states back to the eigenstates of $-i\gamma _{Q_{1}}\gamma _{W_{1}}$. With
this basis transformation method, the projection measurement $P_{\pm
}^{\left( 2\right) }$ can be performed non-destructively.

However, such basis transformation method does not work for the measurements
$P_{\pm }^{\left( 4\right) }$ because they involve eigenstate measurement of
four vortices. Recent work \cite{Bravyi} showed mathematically that $P_{\pm
}^{\left( 4\right) }$ can be realized deterministically using the auxiliary
entangled state $\left\vert \Psi \right\rangle $, for which we provide a
prescription in this Letter, coupled with the braiding operations and the
fusion measurements. Here we refer the mathematical details of this
measurement to Ref. \cite{Bravyi}. Note that the measurement $P_{\pm
}^{\left( 4\right) }$ can be \textit{deterministically} implemented,
although $\left\vert \Psi \right\rangle $ in our scheme can only be
generated with a 50\% success probability. This is because $\left\vert \Psi
\right\rangle $ can be prepared using off-line procedures that are not
involved in the measurement process. In addition, pairs with non-perfect
entanglement can be purified to pairs of nearly perfect entanglement through
off-line purification processes. Therefore, the controlled phase gate $%
\Lambda \left( \sigma _{z}\right) $ can be implemented with a high accuracy
because the remaining processes only involve the braiding operations and the
fusion measurements.

In summary, we proposed and analyzed a scheme to generate entanglement
between two topological vortex qubits in a $p_{x}+ip_{y}$ atomic superfluid
with the assistance of external flying qubits. The entanglement can be
created and purified off-line and therefore, in spite of being a
non-topological process, does not degrade the actual quantum computation
which continues to use the topologically protected braiding operations. We
showed how to test the violation of Bell's inequality using the obtained
entanglement. Finally, we showed how to deterministically implement a set of
universal quantum gates in the chiral $p$-wave superfluid, which had
hitherto remained a major conceptual problem, using the entanglement created
between two topological qubits.

We thank A. Kitaev for helpful discussion. This work is supported by
LPS-NSA and ARO-DTO. \vskip-6mm

\begin{thebibliography}{99}
\bibitem{Kitaev} A. Kitaev, Ann. Phys. \textbf{303}, 2 (2003).

\bibitem{Read} G. Moore and N. Read, Nucl. Phys. B \textbf{360}, 362 (1991).

\bibitem{Nayak} C. Nayak and F. Wilczek, Nucl. Phys. B \textbf{479}, 529
(1996).

\bibitem{Green} N. Read and D. Green, Phys. Rev. B \textbf{61}, 10267 (2000).

\bibitem{Ivanov} D. A. Ivanov, Phys. Rev. Lett. \textbf{86}, 268 (2001); A.
Stern, \textit{et al.}, Phys. Rev. B \textbf{70}, 205338 (2004); M. Stone
and S.-B. Chung, \textit{ibid} \textbf{73}, 014505 (2006); S. Tewari,
\textit{et al.}, cond-mat/0609556.

\bibitem{Tewari} S. Tewari, \textit{et al.}, Phys. Rev. Lett. \textbf{98},
010506 (2007);

\bibitem{Regal} C. A. Regal, \textit{et al.}, Phys. Rev. Lett. \textbf{90},
053201 (2003).

\bibitem{Gurarie} V. Gurarie, \textit{et al.}, Phys. Rev. Lett. \textbf{94},
230403 (2005).

\bibitem{Cheng} C.-H. Cheng and S.-K. Yip, Phys. Rev. Lett. \textbf{95},
070404 (2005).

\bibitem{Bravyi} S. Bravyi, Phys. Rev. A \textbf{73}, 042313 (2006).

\bibitem{Rezayi} N. Read and E. Rezayi, Phys. Rev. B \textbf{59}, 8084
(1999).

\bibitem{Freedman} M. Freedman, \textit{et al.}, Comm. Math. Phys. \textbf{%
227}, 605 (2002).

\bibitem{Bonesteel} N.E. Bonesteel, \textit{et al.}, Phys. Rev. Lett.
\textbf{95}, 140503 (2005).

\bibitem{Walker} M. Freedman, \textit{et al.}, Phys. Rev. B \textbf{73},
245307 (2006).

\bibitem{Sarma} S. Das Sarma, \textit{et al.}, Phys. Rev. Lett. \textbf{94},
166802 (2005); P. Bonderson, \textit{et al.}, \textit{ibid} \textbf{96},
016803 (2006); A. Stern and B.I. Halperin, \textit{ibid} \textbf{96}, 016802
(2006).

\bibitem{Georgiev} L. Georgiev, Phys. Rev. B \textbf{74}, 235112 (2006).

\bibitem{Zoller} D. Jaksch, \textit{et al.}, Phys. Rev. Lett. \textbf{82},
1975 (1999).

\bibitem{Clauser} J.F. Clauser and A. Shimony, Rep. Prog. Phys. \textbf{41},
1981 (1978).

\bibitem{Kitaev2} S. Bravyi and A. Kitaev, Ann. Phys., \textbf{298}, 210
(2002).
\end{thebibliography}

\end{document}